# Helping Computers Understand Geographically-Bound Activity Restrictions

**Marcus Soll, Philipp Naumann**
University of Hamburg
{2soll,2naumann}@
informatik.uni-hamburg.de

**Johannes Schöning, Pavel Samsonov**
Expertise Centre for Digital Media
Hasselt University - tUL - iMinds
<*First.Last*>@uhasselt.be

**Brent Hecht**
Dept. of Comp. Sci. and Engineering
University of Minnesota,
bhecht@cs.umn.edu

**ABSTRACT**
The lack of certain types of geographic data prevents the development of location-aware technologies in a number of important domains. One such type of "unmapped" geographic data is space usage rules (SURs), which are defined as geographically-bound activity restrictions (e.g. "no dogs", "no smoking", "no fishing", "no skateboarding"). Researchers in the area of human-computer interaction have recently begun to develop techniques for the automated mapping of SURs with the aim of supporting activity planning systems (e.g. one-touch "Can I Smoke Here?" apps, SUR-aware vacation planning tools). In this paper, we present a novel SUR mapping technique – *SPtP* – that outperforms state-of-the-art approaches by 30% for one of the most important components of the SUR mapping pipeline: associating a point observation of a SUR (e.g. a 'no smoking' sign) with the corresponding polygon in which the SUR applies (e.g. the nearby park or the entire campus on which the sign is located). This paper also contributes a series of new SUR benchmark datasets to help further research in this area.

**ACM Classification Keywords**
H.5.m. Information interfaces and presentation (e.g., HCI): Miscellaneous;

**INTRODUCTION & MOTIVATION**
In the past decade, location-aware technologies have made the leap from research prototypes to mainstream systems. These technologies are pervasive: they guide people from point A to point B, they help them decide where to eat, they supply them with contextually relevant news and information, and they provide them with many other important services [3,5,16]. However, despite massive increases in the prevalence and diversity of location-aware technologies, many potential location-aware technologies are held back by the lack of data about important geographic phenomena [11,12]. If these phenomena were to be mapped, entirely new classes of location- aware technologies would be enabled.

One such "unmapped" phenomenon that has recently received attention in HCI is that of space usage rules (SURs) [10, 11]. Space usage rules are geographically-bound activity restrictions such as "no smoking", "no fishing", "no swimming", and "no campfires". Despite the key role of these rules in protecting public health, enforcing the law, and maintaining the environment, there is no large dataset of SURs in existence [11]. Indeed, the most extensive SUR dataset available lies in OpenStreetMap's tags (e.g. "smoking=no", "fishing=no") and these tags have extremely limited coverage (e.g. only 7 locations are tagged with "fishing=no" in all of North America [10] and only around 6,000 objects out of the over 3 billion objects in OSM are tagged with "dogs=no, on leash" [4]).

With the aim of reducing the paucity of digital information about SURs, Samsonov et al. [10] recently introduced a method to mine SURs from publicly available geotagged Flickr photos using computer vision techniques. In this paper, the authors demonstrated that their approach is capable of identifying SUR indicators (e.g. "no dog" signs) in the background and foreground of uploaded photos. However, as Samsonov et al. point out, identifying the presence of a SUR indicator at a given latitude and longitude coordinate (e.g. a geotag) is only the first of two challenges associated with SUR mapping. The second challenge – determining the polygon (i.e. area) in which he observed SUR applies – was not the focus of Samsonov et al. [10], with the authors only attempting very straightforward approaches (e.g. selecting the nearest OpenStreetMap (OSM) polygon).

This note presents the first work to rigorously consider the challenge of matching a point SUR observation to the polygon in which the SUR applies, a problem we label the *SUR Association Problem*. The core contribution of this research is a novel technique – *Smart-Point-to-Polygon* (*SPtP*) – whose accuracy on the SUR Association Problem is almost 30% greater than the current state-of-the-art (Samsonov et al.'s work). *SPtP*, which recently won a Germany-wide computer science competition (http://informaticup.gi.de) on SUR mapping, also has the important advantage of being able to function in a wider variety of geographic contexts relative to the state-of-the-art, for instance places with poor OpenStreetMap (OSM) polygon coverage. These improvements are enabled by *SPtP*'s combination of machine learning techniques — in particular ensemble learning and genetic algorithms — with an understanding of spatial reasoning and the built environment.

The research described below also makes two supporting contributions. First, along with this note, we are releasing three new SUR Association Problem evaluation datasets, datasets that are up to 4 times larger than was previously available. Second, although research on SUR mapping has thus far largely occurred within HCI, we believe other areas of computer science can also contribute to this domain. In this vein, we also contribute the first formal definition of the SUR Association Problem, encoding the problem in a fashion that is amenable to a variety of machine learning approaches.

**THE SUR ASSOCIATION PROBLEM**

We can define the SUR Association Problem as mapping a point observation of a SUR (e.g. a "no smoking" sign near a park) $p_{SUR}$ with the corresponding polygon $P_{TARGET}$ to which the SUR applies (e.g. a park to which the sign is referring). Arbitrarily choosing a bounding box around $p_{SUR}$ will not yield useful results, as SURs typically apply to distinct spatial features, as it is outlined in prior work by Samsonov et al. [10]. Thus the SUR Association Problem boils down to selecting the polygon that is most likely to match $P_{TARGET}$ (i.e. the intended area the SUR applies to) from the set of potential candidate polygons $\Omega_P$.

To measure the performance of SUR Association Problem algorithms, we use the polygon intersection ratio (*R*) [9]. *R* operationalizes the assumption that two polygons are more similar when they share more of their areas. More formally, the intersection ratio between a candidate SUR polygon $P_{CANDIDATE} \in \Omega_P$ and the actual SUR polygon $P_{TARGET}$ is:

$$R_{Intersection}(P_{Target}, P_{Cand.}) = \frac{2A\ Intersection(P_{Target}, P_{Cand.})}{A\ P_{Target} + A\ P_{Cand.}}$$

**DATA SOURCES & DATASETS**

We use OpenStreetMap data (as in prior work [10]) to find or create candidate polygons $\Omega_P$ around a given $p_{SUR}$. We also test our algorithms using OpenStreetMap polygons. Below, we describe our data sources, training datasets and test data in more detail.

OSM often contains hundreds or even thousands of geographic entities to which SURs around a $p_{SUR}$ can be associated (e.g. restaurants, schools, churches, stores, parks, airports). Therefore, we limit the set of potential candidate polygons $\Omega_P$ by just considering OSM features in a 500m radius around a $p_{SUR}$. These entities are represented as either points ("nodes") or polygons ("ways"). The polygons can be used as-is as $P_{TARGET}$ candidates for the SUR Association Problem, but the points must first be transformed into polygons. We do so by generating a circular area around each point or "node", treating this new area as a polygon to which a SUR can be associated. The radius of this area differs based on the type of geographic entity under consideration, e.g. a train station gets a larger radius than a restaurant.

To train our algorithm, the authors collected a dataset of 207 geotagged photos of SURs ($p_{SUR}$) with a total of 305 space usage rules and 53 distinct SURs (ranging from "no smoking" to "no motorcycle helmets in the building") mostly in the Hamburg, Germany metropolitan area. In addition, we also manually collected the corresponding $P_{TARGET}$ polygons. We refer to this dataset as *TRAINING207*.

To test our developed algorithm, which we describe in detail below, we used three distinct test data sets with multiple $p_{SUR}$ and their corresponding $P_{TARGET}$ polygons. The first data set, which we refer to as *EVAL96*, was created by university students in eastern Belgium. As part of a class assignment, the students were sent out to find SURs in their neighborhoods and the students collected $p_{SUR}$ with the corresponding polygons $P_{TARGET}$. *EVAL96* contains 96 geotagged photos of SURs with a total of 128 space usage rules and 21 different SURs (ranging from "no alcohol consumption" to "no cellphone use"). The second dataset, *EVAL102*, contains 102 different geotagged SUR photos (containing 150 SURs and 40 distinct types of SUR), also from the Hamburg area (no photos appearing in *EVAL102* also appear in *TRAINING207*). Finally, *EVAL243* consists of data provided by the participants of the InformatiCup. The InformatiCup is a yearly programming challenge organized by the "*Gesellschaft für Informatik*", the national computer science society of Germany. Around 30 different groups participated in this challenge and provided 243 geotagged SUR photos (and their corresponding $P_{TARGET}$). These photos contain 427 space usage rule indicators from all over Germany, with 52 distinct types of indicators.

**THE SPTP ALGORITHM**

Wolpert and Macready [15] describe the "No Free Lunch Theorem" as a general limitation for optimization problems. They show that "the computational cost of finding a solution, averaged over all problems in the class, is the same for any solution" [17]. In the context of the SUR Association Problem, this suggests that if we find a single algorithm that selects good polygons for some types of SURs and SUR locations, there will SURs and SUR locations for which other algorithms will do better. As has been done in prior work [2,7,14], we use ensemble learning [6,8] to address this general challenge for optimization problems. The basic idea is the following: by combining different algorithms that are better than random guessing (referred to as *weak* *classifiers*), we get a single algorithm (referred to as a *strong classifier*) that performs better than any weak classifier alone. The goal is that each time a weak classifier is added, the error rate of the strong classifier is reduced. Another important aspect of ensemble learning is that each of the weak classifiers should be optimally weighted by its performance (Table 1). To calculate these weights, we use a simple genetic algorithm [1] as described below.

**Weak Classifiers**
All of the weak classifiers we developed rate each candidate polygon $P_{CANDIDATE} \in \Omega_P$ in a radius of 500m around a given $p_{SUR}$ with a real number between 100 (very likely to match $P_{TARGET}$) and -100 (very unlikely to match $P_{TARGET}$), with zero being a purely neutral value. The rating of each weak classifier is then multiplied with the classifier's learned weight, and these values are then summed together across all weak classifiers. This process results in each candidate polygon $P_{CANDIDATE} \in \Omega_P$ receiving a score $S(P_{CANDIDATE})$ from the ensemble classifier. The polygon with the highest $S(P_{CANDIDATE})$ is then selected by the classifier as the most likely extent of the SUR indicated in the corresponding point SUR observation $p_{SUR}$ (e.g. a "no smoking" sign). All classifiers were designed and implemented using prior knowledge we gained by analyzing the dataset *TRANING207* with known $P_{TARGET}$ that were collected by the authors and are distinct from the testing data sets used later. We implemented various different classifiers, which can be divided into the six categories that form the sub-sections below. In a second step, a genetic algorithm determined the weights of the classifiers, which is described in more detail following the discussion of the weak classifiers.

*Distance-based Classifiers*
We hypothesized that, in general, $P_{TARGET}$ polygons located close to $p_{SUR}$ are more likely to be those to which the corresponding SUR applies. As such, we implemented 3 distance-based classifiers, each of which rates candidate polygons according to a different distance metric: (1) the distance between $p_{SUR}$ to polygon's centroid, (2) the distance of $p_{SUR}$ to the polygon's closest edge, and (3) the distance of $p_{SUR}$ to polygon's closest vertex. The second algorithm is the algorithm used by Samsonov et al.

*Point in Polygon Classifier*
The "Point in Polygon classifier" is straightforward: it gives a $P_{TARGET}$ a rating of 100 if the candidate polygon's area contains $p_{SUR}$. If not, the candidate is given a rating of -75.

*OpenStreetMap Tag-based classifiers*
As noted above, in addition to spatial entities, OpenStreetMap also contains "tags" for these entities. These tags describe everything from the main function of points and polygons (e.g. "amenity=restaurant", "cuisine=icecream") to their important features (e.g. "wheelchair=yes") to, in rare cases, space usage rules (e.g. "smoking = no", as discussed above). While analyzing the data of *TRANING207* with known $P_{TARGET}$, we noticed that there are some combinations of space usage rules and tags that are more likely to occur than others. For example, "no swimming" is more likely to be associated with a polygon containing the tag "natural=water" than one containing the tag "amenity=restaurant". Based on this observation, we implemented two additional weak classifiers named SUR Description and SUR OSM Mapping, with SUR Description handling one-to-many rules (e.g. "no swimming" could be associated with a beach, pool, lake, etc.) and SUR OSM Mapping handling one-to-one rules (e.g. "no motorcycle helmets inside" is frequently a rule for banks in Europe, but only banks). *SPtP* has 16 SUR Description rules and 33 SUR OSM Mapping rules in its current version, all of which were manually developed. Full details describing all rules can be found in the source for *SPtP* (see below).

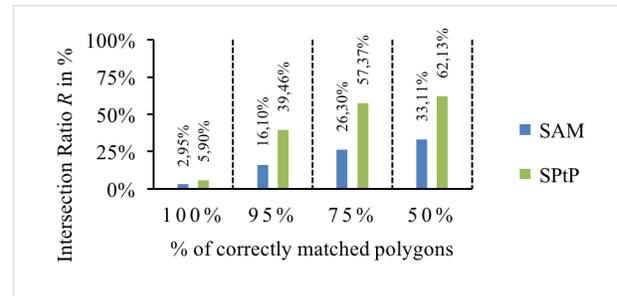

**Figure 1:** Comparison of the number and percentages of correctly identified polygons by Samsonov et al. (SAM) and SPtP (SPtP) averaged over all three evaluation datasets. The graph shows the minimum intersection ratio for a polygon to be eligible to be "correct".

*Orientation-based classifier*
Although there are other possible approaches (see below), all prior work involving SURs has used geotagged photos as the point SUR observations $p_{SUR}$ (and this is true of all of our datasets). When storing a geotagged photo, some cameras automatically encode the camera's orientation (rotation) into the photo's Exif metadata. We developed a classifier that is able to leverage this orientation information (when it is available) by giving polygons that are in the line of sight of the camera higher ratings.

*Computer Vision-based classifier*
In addition to leveraging information from Exif metadata, we also developed a classifier that uses the images themselves. This classifier determines whether a photo was taken outside or inside using an approach based on Szummer et al [13]. Once the inside/outside property of the image was determined, we compared this with a manually developed mapping from (SUR type, type of space {inside, outside}) tuples (e.g. "no smoking" and inside) to the type of polygon to which this tuple is likely to apply (e.g. "amenity=building" in the case of "no smoking" and inside). Polygons that matched this mapping were given a high rating, and those that did not were given a low rating.

*Genetic algorithm*

As noted above, we used genetic algorithm to determine the weights (influence) to assign to each of our weak classifiers. The first population of weights was assigned to random values, and descendants of this population were determined using standard genetic algorithm approaches. Fitness of each descendant was determined by comparing the corresponding strong classifier's prediction for each $p_{SUR}$ in *TRAINING207* against its human-labeled $P_{TARGET}$ (using the polygon intersection ratio $R$).

**EVALUATION & RESULTS**

As described above in the datasets section, we used three different datasets for the evaluation of *SPtP* and the intersection ratio (R) as our evaluation metric. In addition, we have counted the number of correct associations of $P_{SUR}$ to $P_{TARGET}$, with "correct" defined by different intersection ratio values ranging from at least 5% overlap to at least 50% overlap (Table 1). The evaluation was done based on OSM data downloaded in June 2015.

As seen in Table 1 and Figure 1, our approach exceeds the approach of Samsonov et al. for all definitions of correct and in all datasets. On all three datasets, our approach suggests $P_{TARGET}$ polygons with an average intersection ratio of about 60%. It can also be seen that the approach of Samsonov et al. has problems with larger and more diverse datasets such as EVAL102 and EVAL243, whereas the performance of our approach stays around 60 % and is even the highest for the most complex dataset EVAL243. Not surprisingly the most matches with R = 100% in our algorithm (an exact match of a $P_{CANDIDATE}$ with a $P_{TARGET}$) was found in the *EVAL102*, the dataset that had the same regional scope as *TRAINING207* (26 out of 102 had R = 100, as compared to 23 out of 102 for Samsonov et al.). In EVAL96 and EVAL243, *SPtP* achieved R=100% for 0 polygons and R=75% for 143 polygons, respectively (as compared to 0 and 54 polygons for Samsonov et al).

Unpacking the performance of *SPtP*, we found that certain weak classifiers were more effective than others (table 1). For example, both tag-based classifiers had higher weights (4.3 and 3.5) compared to the weights of the distance-based classifier (0.5, 2.8, 0.5). Interestingly the computer vision-based classifier also had a relative high weight (1.0) compared to the distance-based classifier.

**CONCLUSION & FUTURE WORK**

In this paper, we have presented a new technique, *SPtP*, that by using ensemble learning and genetic algorithms improves our ability to associate point space usage rule observations with their corresponding area of application. In addition, we have contributed three new datasets (*TRAINING207*, *EVAL96*, *EVAL102, EVAL243*) that can be used for the training and evaluation of future SUR Association Problem approaches. All of the datasets, as well as our *SPtP* code, are available on online at https://github.com/Top-Ranger/SPtP, and we invite other researchers to attempt to replicate and extend our approach.

|  | EVAL96 | EVAL102 | EVAL243 |
|---|---|---|---|
| **Dist. Centroid** (identical to SAM [10]) | **39.3 %** | **33.3 %** | **30.9 %** |
| Dist. Closest Edge | 37.5 % | 37.7 % | 39.7 % |
| Dist. Closest Vertex | 36.9 % | 31.6 % | 34.6 % |
| SUR Description | 5.1 % | 2.2 % | 4.8 % |
| SUR OSM Mapping | 5.3 % | 4.5 % | 2.9 % |
| Orientation | 6.5 % | 2.1 % | 5.8 % |
| Computer Vision | 2.7 % | 0.6 % | 2.7 % |
| SPtP – Equal Weights | 54.7 % | 45.3 % | 45.7 % |
| **SPtP – Weights with Genetic Algorithm** | **62.2 %** | **60.2 %** | **63.2 %** |

**Table 1:** Overview of the individual performance of the different weak classifiers as well as the average intersection ratio of the approaches of Samsonov et al. (SAM) and our technique SPtP. The Samsonov et al. approach is identical to our first weak classifier cf. [10] (analysis based on OSM data from Nov. 2015)

Although we outperformed the SUR Association Problem state-of-the-art, there is much future work to do: 1) There are likely more weak classifiers that can be developed, and the weak classifiers we did include can likely be improved. One promising avenue involves incorporating non-OSM data (e.g. algorithms that operate on remotely sensed imagery). 2) SUR mapping is not the only problem that can be supported by "smart point-to-polygon" approaches. For instance, *SPtP* may be useful for associating photos with their subject, rather than their specific geotag. 3) Some SUR observations may be associated with multiple polygons. Thanks to OpenStreetMap's *relations*, which link together related polygons (e.g. islands and their corresponding country), *SPtP* supports these situations where relations have been encoded. However, in many places, relations are not common. Future work should examine an extension of the SUR Association Problem in which a SUR can be associated with several polygons. Finally, the increased accuracy of *SPtP* helps to open up new possibilities for SUR mapping more generally. For instance, we have created a working prototype of an app to support the explicit crowdsourcing of SUR mapping. The app allows contributors to snap a photo of a SUR sign (e.g. "no dogs") and select the polygon in which the corresponding SUR applies. By incorporating *SPtP* into this app and automatically suggesting polygons in a ranked fashion, we will decrease the effort associated with each new SUR observation and hopefully increase the number of SURs that get mapped.

**ACKNOWLEDGMENTS**

We would like to thank the organizers, the jury and all participants of the InformatiCup 2015. The work was also supported by the following grants: A Google Research Faculty Award, BOF R-5209, NSF IIS-1526988, and FWO K207615N.

*Note: This version of the paper contains a fix for a reference issue that appeared in the original version.*